\documentclass{aastex}%
\usepackage{amssymb}%
\usepackage{amsmath}%
\usepackage{amsfonts}%
\usepackage{graphicx}
\begin{document}
{\LARGE Streaming cold cosmic ray back-reaction and thermal instabilities}

\bigskip

\ \ \ \ \ \ \ \ \ \ \ \ \ \ \ \ \ \ \ \ \ {\large Anatoly K. Nekrasov}$^{1}%
${\large \ and Mohsen Shadmehri}$^{2}$

$^{1}$ Institute of Physics of the Earth, Russian Academy of Sciences, 123995
Moscow, Russia;

anekrasov@ifz.ru, nekrasov.anatoly@gmail.com

$^{2}$ Department of Physics, Golestan University, Basij Square, Gorgan, Iran;

m.shadmehri@gu.ac.ir

\bigskip

\ \ \ \ \ \ \ \ \ \ \ \ \ \ \ \ \ \ \ \ \ \ \ \ \ \ \ \ \ \ \ \ \ \ \ \ \ \ \ \ \ \ \ \ \ \ \ \ \ \ ABSTRACT

We investigate the streaming and thermal instabilities of the electron-ion
plasma with homogeneous cold cosmic rays drifting perpendicular to the
background magnetic field in the multi-fluid approach. One-dimensional
perturbations along the magnetic field are considered. The induced return
current of the background plasma and back-reaction of cosmic rays are taken
into account. It is shown that the cosmic ray back-reaction results in the
streaming instability having considerably larger growth rates than that due to
the return current of the background plasma. This increase is by a factor of
the square root of the ratio of the background plasma mass density to the
cosmic ray mass density. The maximal growth rates and corresponding wave
numbers are found. The thermal instability is shown to be not subject to the
action of cosmic rays in the model under consideration. The dispersion
relation for the thermal instability includes ion inertia. In the limit of
fast thermal energy exchange between electrons and ions, the isobaric and
isochoric growth rates are derived. The results obtained can be useful for the
investigation of the electron-ion\textbf{\ }astrophysical objects\textbf{\ }%
such as galaxy clusters including the dynamics of streaming cosmic rays.

\bigskip


\bigskip

\section{INTRODUCTION}

There is a growing interest towards understanding of interactions of cosmic
rays with plasma systems in astrophysics and their possible effects.
Irrespective of various mechanisms which are proposed for the generation of
such high energy particles, cosmic rays may interact with the existing
turbulent motions in a plasma or even may excite them. In order to study
cosmic rays, a particle description is needed, although the fluid
approximation is used as well for simplicity. Cosmic rays may further induce
ionization which may dramatically change the physical properties of a system.
For example, the ionization by cosmic rays has a vital role in star formation
near the Galactic center (e.g., Yusef-Zadeh et al. 2007) or in the dead zone
of protoplanetary disks (Gammie 1996). On the other hand, the heating rate is
enhanced because of cosmic rays and this important effect has been studied in
the context of structure formation in the interstellar medium (ISM) via
thermal instability (e.g., Goldsmith et al. 1969; Field et al. 1969).

Another contribution of cosmic rays to the dynamical evolution of the system
is their pressure. Many authors studied the dynamical role of cosmic rays in
structure formation at large scales by Parker instability (Parker 1966;
Kuwabara \& Ko 2006), magnetorotational instability (Khajenabi 2012) and even
galactic winds or outflows (e.g., Everett et al. 2008). Recently, Wagner et
al. (2005) and Shadmehri (2009) extended the classical thermal instability
(Field 1965) to include cosmic rays. The thermal instability has been used to
explain existence of structures not only in the ISM but also at the very large
scales like in the intracluster medium (ICM). This instability is assumed as a
possible mechanism for producing molecular filaments (Sharma et al. 2010) seen
in galaxy clusters with short ($\lesssim$1 Gyr) cooling times (e.g., Conselice
et al. 2001; Salom\'{e} et al. 2006; Cavagnolo et al. 2008; O'Dea et al.
2008). Recent linear analysis of thermal instability with cosmic ray pressure
shows that the instability is suppressed (e.g., Shadmehri 2009). Cosmic rays
has been included by Shadmehri (2009) (see also Sharma et al. 2010) in the
framework of the magnetohydrodynamic equations as a second fluid having the
velocity of the thermal plasma. Numerical analysis shows that the cosmic ray
pressure can play an important role in the dynamics of cold filaments making
them much more elongated along the magnetic field lines than the Field length.
This is consistent with observations as well (Sharma et al. 2010; see also
Snodin et al. 2006). Also, including cosmic rays is required to explain the
atomic and molecular lines observed in filaments in clusters of galaxies
(Ferland et al. 2009).

However, there is another important effect of cosmic rays which has not been
considered in the context of thermal instability, namely the presence of
streaming cosmic rays. These particles are charged and their\textbf{\ }drift
motion induces a current. It has been assumed that at some distance from the
shock cosmic rays are mainly positively charged particles (Riquelme \&
Spitkovsky 2009). The cosmic ray drift driving a constant current results in
arising of the return current provided by the background plasma (e.g., Bell
2004, 2005; Riquelme \& Spitkovsky 2009). The possible role of this effect in
the generation of thermal instability needs to be considered. There is also
another important issue like the amplification of magnetic fields. The
classical cyclotron resonant instability has been proposed long time ago to
explain this process (Kulsrud \& Pearce 1969). However, this mechanism has
turned out to be unable to provide sufficient energy in the shock upstream
plasma. In order to resolve this problem, just recently a new non-resonant
instability has been introduced that may provide a much higher energy (Bell
2004). This instability, which is known as the Bell instability, has also been
confirmed by non-linear numerical simulations (e.g., Riquelme \& Spitkovsky
2009). Subsequent works extended this instability into various directions by
considering various physical factors (e.g., Reville et al. 2007; Reville \&
Bell 2012). However, all these works, except the paper by Bell (2005), have
been restricted to the cosmic ray drift speed parallel to the initial magnetic
field and correspondingly to unmagnetized cosmic rays, where Larmor radii of
cosmic rays defined by longitudinal velocities are much larger than the length
scale of interest (see also Zweibel 2003). In the case of magnetized cosmic
rays, whose Larmor radii are smaller than the typical length scales of the
system, the cosmic ray current can be perpendicular to the initial magnetic
field as has been demonstrated by Riquelme \& Spitkovsky (2010). Riquelme \&
Spitkovsky (2010) studied this perpendicular current-driven instability in the
linear regime and also numerically and compared their growth rate with the
cosmic ray current-driven instability (CRCD) by Bell (2004). The growth rates
and length scales in both cases were similar. But the authors have not
included the cosmic ray back-reaction analytically.

Depending on the magnetic field strength and the cosmic ray flux, the cosmic
ray thermal pressure effect may become important and modify the maximum growth
rate of the CRCD (for a detailed discussion, see Zweibel \& Everett 2010). The
thermal instability in galaxy clusters in the multi-fluid approach has been
considered by Nekrasov (2011, 2012). The related effects by cosmic rays were
not included in these papers. Here, we take into account streaming cold cosmic
rays. We consider a geometry in which homogeneous cosmic rays drift across the
background magnetic field and perturbations arise along the latter. Such a
geometry is analogous to that treated by Riquelme \& Spitkovsky (2010).
However, we include the cosmic ray back-reaction. We also take into account
the plasma return current (e.g., Bell 2004, 2005; Riquelme \& Spitkovsky 2009,
2010). For simplicity, we ignore the action of gravity here\textbf{\ }(Sharma
et al. 2010). The effects of the gravitational field have been investigated in
detail in multi-fluid approach in papers by Nekrasov \& Shadmehri (2010,
2011). Thus, our present study extends previous analytical studies by
considering not only the thermal effects but the currents driven by cosmic
rays and their back-reaction.

The paper is organized in the following manner. Section 2 contains the
fundamental equations for plasma, cosmic rays, and electromagnetic fields used
in this paper. Equilibrium state is discussed in Section 3. In Sections 4 and
5, the perturbed velocities of the ions and electrons and perturbed plasma
current are given, respectively. Corresponding results obtained for cosmic
rays are contained in Sections 6 and 7. The total perturbed current is given
in Section 8. Wave equations are found in Section 9. Dispersion relation
including the plasma return current, cosmic ray back-reaction, and terms
describing the thermal instability is derived in Section 10. In Section 11, a
discussion of important results obtained is provided. Possible astrophysical
implications are given in Section 12. Conclusive remarks are summarized in
Section 13.

\bigskip

\section{BASIC\ EQUATIONS FOR\ PLASMA\ AND\ COSMIC\ RAYS}

The fundamental equations for a plasma that we consider here are the
following:%
\begin{equation}
\frac{\partial\mathbf{v}_{j}}{\partial t}+\mathbf{v}_{j}\cdot\mathbf{\nabla
v}_{j}=-\frac{\mathbf{\nabla}p_{j}}{m_{j}n_{j}}+\frac{q_{j}}{m_{j}}%
\mathbf{E+}\frac{q_{j}}{m_{j}c}\mathbf{v}_{j}\times\mathbf{B},
\end{equation}
the equation of motion,%
\begin{equation}
\frac{\partial n_{j}}{\partial t}+\mathbf{\nabla}\cdot n_{j}\mathbf{v}_{j}=0,
\end{equation}
the continuity equation,%
\begin{equation}
\frac{\partial T_{i}}{\partial t}+\mathbf{v}_{i}\cdot\mathbf{\nabla}%
T_{i}+\left(  \gamma-1\right)  T_{i}\mathbf{\nabla}\cdot\mathbf{v}%
_{i}=-\left(  \gamma-1\right)  \frac{1}{n_{i}}%
\mathcal{L}%
_{i}\left(  n_{i},T_{i}\right)  +\nu_{ie}^{\varepsilon}\left(  n_{e}%
,T_{e}\right)  \left(  T_{e}-T_{i}\right)
\end{equation}
and%
\begin{align}
\frac{\partial T_{e}}{\partial t}+\mathbf{v}_{e}\cdot\mathbf{\nabla}%
T_{e}+\left(  \gamma-1\right)  T_{e}\mathbf{\nabla}\cdot\mathbf{v}_{e}  &
=-\left(  \gamma-1\right)  \frac{1}{n_{e}}\mathbf{\nabla\cdot q}_{e}-\left(
\gamma-1\right)  \frac{1}{n_{e}}%
\mathcal{L}%
_{e}\left(  n_{e},T_{e}\right) \\
& -\nu_{ei}^{\varepsilon}\left(  n_{i},T_{e}\right)  \left(  T_{e}%
-T_{i}\right) \nonumber
\end{align}
are the temperature equations for ions and electrons. In Equations (1) and
(2), the subscript $j=i,e$ denotes the ions and electrons, respectively.
Notations in Equations (1)-(4) are the following: $q_{j}$ and $m_{j}$ are the
charge and mass of species $j$, $\mathbf{v}_{j}$ is the hydrodynamic velocity,
$n_{j}$ is the number density, $p_{j}=n_{j}T_{j}$ is the thermal pressure,
$T_{j}$ is the temperature, $\nu_{ie}^{\varepsilon}(n_{e},T_{e})$ ($\nu
_{ei}^{\varepsilon}\left(  n_{i},T_{e}\right)  $) is the frequency of thermal
energy exchange between ions (electrons) and electrons (ions) being $\nu
_{ie}^{\varepsilon}(n_{e},T_{e})=2\nu_{ie}$, where $\nu_{ie}$ is the collision
frequency of ions with electrons (Braginskii 1965), $n_{i}\nu_{ie}%
^{\varepsilon}\left(  n_{e},T_{e}\right)  =n_{e}\nu_{ei}^{\varepsilon}\left(
n_{i},T_{e}\right)  $, $\gamma$ is the ratio of the specific heats,
$\mathbf{E}$\textbf{\ }and $\mathbf{B}$ are the electric and magnetic fields,
and $c$ is the speed of light in a vacuum. Here, for simplicity, we do not
take into account collisions between the ions and electrons in the momentum
equation. This effect for the thermal instability has been treated by Nekrasov
(2011, 2012), where, in particular, conditions allowing to neglect the
collisions have been found. However, the thermal exchange should be included
because its time scale compares with the dynamical time. The value
$\mathbf{q}_{e}$ in Equation (4) is the electron heat flux (Braginskii 1965).
In a weakly collisional plasma, which is considered here, the electron Larmor
radius is much smaller than the electron collisional mean free path. In this
case, the electron thermal flux is mainly directed along the magnetic field,
\begin{equation}
\mathbf{q}_{e}=-\chi_{e}\mathbf{b}\left(  \mathbf{b\cdot\nabla}\right)  T_{e},
\end{equation}
where $\chi_{e}$ is the electron thermal conductivity coefficient and
$\mathbf{b=B/}B$ is the unit vector along the magnetic field. We only take
into account the electron thermal flux given by Equation (5) because the
longitudinal ion thermal conductivity is considerably smaller (Braginskii
1965). We also assume that the thermal flux in the equilibrium is absent. The
cooling and heating of plasma species in Equations (3) and (4) are described
by function $%
\mathcal{L}%
_{j}(n_{j},T_{j})=n_{j}^{2}\Lambda_{j}\left(  T_{j}\right)  -n_{j}\Gamma_{j}$,
where $\Lambda_{j}$ and $\Gamma_{j}$ are the cooling and heating functions,
respectively. The form of this function is somewhat different from the usually
used cooling-heating function $\pounds $ (Field 1965). Both functions are
connected to each other via the equality $%
\mathcal{L}%
_{j}\left(  n_{j},T_{j}\right)  =m_{j}n_{j}\pounds _{j}$. Our choice is
analogous to those of Begelman \& Zweibel (1994), Bogdanovi\'{c} et al.
(2009), Parrish et al. (2009). The function $\Lambda_{j}\left(  T_{j}\right)
$ can be found, for example, in Tozzi \& Norman (2001).

Equations for relativistic cosmic rays we take in the form (e.g. Lontano et
al. 2002)
\begin{equation}
\frac{\partial\left(  R_{cr}\mathbf{p}_{cr}\right)  }{\partial t}%
+\mathbf{v}_{cr}\cdot\mathbf{\nabla}\left(  R_{cr}\mathbf{p}_{cr}\right)
=-\frac{\mathbf{\nabla}p_{cr}}{n_{cr}}+q_{cr}\left(  \mathbf{E+}\frac{1}%
{c}\mathbf{v}_{cr}\times\mathbf{B}\right)  ,
\end{equation}%
\begin{equation}
\left(  \frac{\partial}{\partial t}+\mathbf{v}_{cr}\cdot\mathbf{\nabla
}\right)  \left(  \frac{p_{cr}\gamma_{cr}^{\Gamma_{cr}}}{n_{cr}^{\Gamma_{cr}}%
}\right)  =0,
\end{equation}
where%
\begin{equation}
R_{cr}=1+\frac{\Gamma_{cr}}{\Gamma_{cr}-1}\frac{T_{cr}}{m_{cr}c^{2}}.
\end{equation}
In these equations, $\mathbf{p}_{cr}=\gamma_{cr}m_{cr}\mathbf{v}_{cr}$ is the
momentum of a cosmic ray particle having the rest mass $m_{cr}$ and velocity
$\mathbf{v}_{cr}$, $q_{cr}$ is the charge, $p_{cr}=\gamma_{cr}^{-1}%
n_{cr}T_{cr}$ is the kinetic pressure, $n_{cr}$ is the number density in the
laboratory frame, $\Gamma_{cr}$ is the adiabatic index, $\gamma_{cr}=\left(
1-\mathbf{v}_{cr}^{2}/c^{2}\right)  ^{-1/2}$ is the relativistic factor. The
continuity equation is the same as Equation (2) for $j=cr$. Equation (8) can
be used for both cold nonrelativistic, $T_{cr}\ll$ $m_{cr}c^{2}$, and hot
relativistic, $T_{cr}\gg$ $m_{cr}c^{2}$, cosmic rays. In the first (second)
case, we have $\Gamma_{cr}=5/3$ ($4/3$) (Lontano et al. 2002) . The general
form of the value $R_{cr}$ applying at any relations between $T_{cr}$ and
$m_{cr}c^{2}$, can be found e.g. in Toepfer (1971) and Dzhavakhishvili and
Tsintsadze (1973).

Equations (1)-(4), (6), and (7) are solved together with Maxwell's equations
\begin{equation}
\mathbf{\nabla\times E=-}\frac{1}{c}\frac{\partial\mathbf{B}}{\partial t}%
\end{equation}
and
\begin{equation}
\mathbf{\nabla\times B}=\frac{4\pi}{c}\mathbf{j+}\frac{1}{c}\frac
{\partial\mathbf{E}}{\partial t},
\end{equation}
where $\mathbf{j=j}_{pl}+\mathbf{j}_{cr}=\sum_{j}q_{j}n_{j}\mathbf{v}%
_{j}+\mathbf{j}_{cr}$. Below, we consider at first an equilibrium state in
which there is a stationary cosmic ray current.

\bigskip

\section{EQUILIBRIUM STATE}

We will consider a uniform plasma embedded in an uniform magnetic field
$\mathbf{B}_{0}$ (the subscript $0$ here and below denotes background
parameters) directed along the $z$-axis. We assume that the plasma\textbf{\ }%
is in equilibrium and penetrated by the uniform beam of cosmic rays having the
uniform streaming velocity $\mathbf{u}_{cr}$ along the $y$-axis. It is
reasonable to suppose that in such uniform model the magnetic field due to
cosmic rays will be absent. This picture is analogous to the consideration of
the gravitational potential in the equilibrium state in an infinite uniform
medium having a constant mass density. Then we obtain from Equation (10)
\begin{equation}
\sum_{j}q_{j}n_{j0}\mathbf{v}_{j0}+\mathbf{j}_{cr0}\mathbf{+}\frac{1}{4\pi
}\frac{\partial\mathbf{E}_{0}}{\partial t}=0.
\end{equation}
From Equation (1), we easily find in the equilibrium state%
\begin{equation}
\mathbf{v}_{e0}=c\frac{\mathbf{E}_{0}\times\mathbf{B}_{0}}{B_{0}^{2}%
},\mathbf{v}_{i0}=\mathbf{v}_{e0}+\frac{c}{\omega_{ci}B_{0}}\frac
{\partial\mathbf{E}_{0}}{\partial t},v_{e0z}=v_{i0z}=0,
\end{equation}
where we have assumed that $\partial/\partial t\ll\omega_{cj}$, $\omega_{cj}$
$=q_{j}B_{0}/m_{j}c$ is the cyclotron frequency. Analogously, we obtain from
Equation (6) under condition $R_{cr}\gamma_{cr}\partial/\partial t\ll
\omega_{ccr}$ ($\omega_{ccr}=q_{cr}B_{0}/m_{cr}c$)
\begin{equation}
\mathbf{v}_{cr0}=\mathbf{v}_{e0}+\mathbf{u}_{cr}.
\end{equation}
In Equation (13), we have neglected the polarizational drift of cosmic rays
(the second term on the right hand-side for the ion velocity in Equation
(12)). This can be done due to condition $n_{i0}\gg n_{cr0}$ (see below),
which is always satisfied. Using Equations (12) and (13), we will find the
current $\mathbf{j}_{0}$%
\begin{equation}
\mathbf{j}_{0}=\frac{q_{i}n_{i0}c}{\omega_{ci}B_{0}}\frac{\partial
\mathbf{E}_{0}}{\partial t}+q_{cr}n_{cr0}\mathbf{u}_{cr},
\end{equation}
where we have taking into account the condition of quasi-neutrality
$q_{i}n_{i0}+q_{e}n_{e0}+q_{cr}n_{cr0}=0$. The number density $n_{cr}$ is the
one in the laboratory frame. We note that this condition is satisfied in
astrophysical plasmas due to cosmic ray charge neutralization from the
background environment (Alfv\'{e}n 1939). Substituting Equation (14) into
Equation (11), we obtain%
\[
\left(  \frac{c^{2}}{c_{Ai}^{2}}\mathbf{+}1\right)  \frac{\partial
\mathbf{E}_{0}}{\partial t}+4\pi q_{cr}n_{cr0}\mathbf{u}_{cr}=0,
\]
where $c_{Ai}=\left(  B_{0}^{2}/4\pi m_{i}n_{i0}\right)  ^{1/2}$ is the ion
Alfv\'{e}n velocity. Usually, the inequality $c\gg c_{Ai}$ is satisfied. Thus,
the induced electric field $\mathbf{E}_{0}$ is determined by equation
\begin{equation}
\frac{\partial\mathbf{E}_{0}}{\partial t}=-4\pi q_{cr}n_{cr0}\mathbf{u}%
_{cr}\frac{c_{Ai}^{2}}{c^{2}}.
\end{equation}
We note that this equation has been given in the paper by Riquelme \&
Spitkovsky (2010). Substituting Equation (15) into Equation (12) for ions, we
find the return plasma current%
\begin{equation}
\mathbf{j}_{ret}=q_{i}n_{i0}\left(  \mathbf{v}_{i0}-\mathbf{v}_{e0}\right)
=-q_{cr}n_{cr0}\mathbf{u}_{cr}=-\mathbf{j}_{cr0},
\end{equation}
which\ is equal to the cosmic ray current and has an opposite direction. The
induced plasma current drift velocity $\mathbf{u}_{pl}=\mathbf{v}%
_{i0}-\mathbf{v}_{e0}$ is equal to $\mathbf{u}_{pl}=-\left(  q_{cr}%
n_{cr0}/q_{i}n_{i0}\right)  \mathbf{u}_{cr}$. Using Equation (15), we see that
the polarizational cosmic ray drift velocity $\left(  R_{cr}\gamma
_{cr}c/\omega_{cr}B_{0}\right)  \partial\mathbf{E}_{0}/\partial t$ can be
neglected in comparison with $\mathbf{u}_{cr}$ under condition $m_{i}n_{i0}\gg
R_{cr}\gamma_{cr}m_{cr}n_{cr0}$. If cosmic rays are not too relativistic, this
condition is satisfied. The drift velocity $\mathbf{u}_{pl}$ will be taken
into account at the consideration of thermal instability which is provided below.

We will here consider the case in which background temperatures of the
electrons and ions are equal between\textbf{\ }each other, $T_{e0}%
=T_{i0}=T_{0}$. However, for convenience to follow the symmetric contribution
of the ions and electrons, we keep in general calculations different
temperatures. In this case, the thermal equations in equilibrium are given by%

\begin{equation}%
\mathcal{L}%
_{i}\left(  n_{i0},T_{i0}\right)  =%
\mathcal{L}%
_{e}\left(  n_{e0},T_{e0}\right)  =0.
\end{equation}

\bigskip

\section{PERTURBED\ VELOCITIES\ OF\ IONS\ AND\ ELECTRONS}

Here, we will investigate one-dimensional perturbations depending on the
$z$-coordinate. Equations for perturbed velocities of ions and electrons are
given in the Appendix A. We consider Equation (A3) under condition
$\omega_{cj}^{2}\gg\partial^{2}/\partial t^{2}$. Then, the transverse
velocities are given by%

\begin{align}
v_{j1x}  & =\frac{q_{j}}{m_{j}\omega_{cj}}E_{1y}+\frac{q_{j}}{m_{j}\omega
_{cj}^{2}}\frac{\partial E_{1x}}{\partial t}-\frac{q_{j}}{m_{j}\omega_{cj}%
^{3}}\frac{\partial^{2}E_{1y}}{\partial t^{2}},\\
v_{j1y}  & =-\frac{q_{j}}{m_{j}\omega_{cj}}E_{1x}+\frac{q_{j}}{m_{j}%
\omega_{cj}^{2}}\frac{\partial E_{1y}}{\partial t}+\frac{q_{j}}{m_{j}%
\omega_{cj}^{3}}\frac{\partial^{2}E_{1x}}{\partial t^{2}},\nonumber
\end{align}
where Equation (A2) has been used. We also have taken into account that
$B_{1z}=0$ (see Equation (9)). The longitudinal velocities $v_{i1z}$ and
$v_{e1z}$ are given by Equation (A13)%
\begin{equation}
Lv_{i1z}=H_{i1},Lv_{e1z}=H_{e1},
\end{equation}
where the values $L$, $H_{i1}$, and $H_{e1}$ are defined by Equations (A9),
(A12), and (A14)-(A20).

\bigskip

\section{PERTURBED PLASMA CURRENT}

It is known that contribution of streaming flows to the dispersion relation is
due to the difference of their velocities (e.g., Nekrasov 2008, 2009a, 2009b,
2009c). Therefore, we can not take into account the contribution of the
electric drift $\mathbf{v}_{e0}$. We also could choose the appropriate frame
of reference and make the electric drift initially zero (Riquelme \&
Spitkovsky 2010).

The linear components of the plasma current $\mathbf{j}_{pl1}=$ $\sum_{j}%
q_{j}n_{j0}\mathbf{v}_{j1}+$ $\sum_{j}q_{j}n_{j1}\mathbf{v}_{j0}$ have the
form%
\begin{align}
4\pi j_{pl1x}  & =\left(  \sum_{j}\frac{\omega_{pj}^{2}}{\omega_{cj}}\right)
E_{1y}+\frac{\omega_{pi}^{2}}{\omega_{ci}^{2}}\frac{\partial E_{1x}}{\partial
t}-\frac{\omega_{pi}^{2}}{\omega_{ci}^{3}}\frac{\partial^{2}E_{1y}}{\partial
t^{2}},\\
4\pi j_{pl1y}  & =-\left(  \sum_{j}\frac{\omega_{pj}^{2}}{\omega_{cj}}\right)
E_{1x}+\frac{\omega_{pi}^{2}}{\omega_{ci}^{2}}\frac{\partial E_{1y}}{\partial
t}+\frac{\omega_{pi}^{2}}{\omega_{ci}^{3}}\frac{\partial^{2}E_{1x}}{\partial
t^{2}}\nonumber\\
& -4\pi q_{i}n_{i0}u_{pl}\frac{1}{L}\left(  \frac{\partial}{\partial
t}\right)  ^{-1}\frac{\partial H_{i1}}{\partial z},\nonumber\\
4\pi j_{pl1z}  & =4\pi q_{i}n_{i0}\frac{H_{i1}}{L}+4\pi q_{e}n_{e0}%
\frac{H_{e1}}{L},\nonumber
\end{align}
where $\omega_{pj}=\left(  4\pi n_{j0}q_{j}^{2}/m_{j}\right)  ^{1/2}$ is the
plasma frequency and $u_{pl}=v_{i0y}$. When deriving Equation (20), we have
used Equations (18) and (19) and the continuity equation.

\bigskip

\section{PERTURBED VELOCITY OF COSMIC RAYS}

We assume cold, nonrelativistic, $T_{cr}\ll m_{cr}c^{2}$, cosmic rays here for
which $\Gamma_{cr}=5/3$ (Lontano et al. 2002). In this case, the value
$R_{cr}$ is equal to the unity, $R_{cr}=1$ (see Equation (8)). The interaction
of cosmic rays with the background plasma can be considered without taking
into account the cosmic ray back-reaction and including the latter. If cosmic
rays are stationary in both equilibrium and perturbations, then we have to
take the quasi-neutrality condition in the background state in the form
$q_{i}n_{i0}+q_{e}n_{e0}=0$ to avoid the appearance of the current due to the
electric drift. When the cosmic ray back-reaction is involved, the condition
of quasi-neutrality becomes $q_{i}n_{i0}+q_{e}n_{e0}+q_{cr}n_{cr0}=0$. The
last condition have been used in Section 3. We here consider the back-reaction
of cosmic rays also in perturbations. We assume that the condition
$\omega_{ccr}^{2}\gg\gamma_{cr0}^{4}\partial^{2}/\partial t^{2}$ is satisfied.
For ultrarelativistic cold cosmic rays ($\gamma_{cr}\gg1)$, the last condition
can be violated. Then the velocities $v_{cr1x,y}$ given by Equation (B3) are
the following:%
\begin{align}
v_{cr1x}  & =\frac{q_{cr}}{m_{cr}\omega_{ccr}}E_{1y}+\frac{q_{cr}\gamma
_{cr0}^{3}}{m_{cr}\omega_{ccr}^{2}}\frac{\partial E_{1x}}{\partial t}%
-\frac{q_{cr}\gamma_{cr0}^{4}}{m_{cr}\omega_{ccr}^{3}}\frac{\partial^{2}%
E_{1y}}{\partial t^{2}},\\
v_{cr1y}  & =-\frac{q_{cr}}{m_{cr}\omega_{ccr}}E_{1x}+\frac{q_{cr}\gamma
_{cr0}}{m_{cr}\omega_{ccr}^{2}}\frac{\partial E_{1y}}{\partial t}+\frac
{q_{cr}\gamma_{cr0}^{4}}{m_{cr}\omega_{ccr}^{3}}\frac{\partial^{2}E_{1x}%
}{\partial t^{2}},\nonumber
\end{align}
where we have used Equation (B2).

The longitudinal velocity $v_{cr1z}$ is defined by Equations (B6) and (B7)%
\begin{equation}
L_{cr}v_{cr1z}=H_{cr1}.
\end{equation}
Substituting Equation (21) for $v_{cr1y}$ into Equation (B7) for $H_{cr1}$ and
using Equation (B2), we obtain%
\begin{equation}
H_{cr1}=\frac{q_{cr}}{m_{cr}\omega_{ccr}}c_{scr}^{2}\gamma_{cr0}^{2}%
\frac{u_{cr}}{c^{2}}\frac{\partial^{2}}{\partial z\partial t}\left(
-E_{1x}+\frac{\gamma_{cr0}}{\omega_{ccr}}\frac{\partial E_{1y}}{\partial
t}\right)  +\frac{q_{cr}}{m_{cr}}\frac{\partial}{\partial t}\left(
E_{1z}-\frac{u_{cr}}{c}B_{1x}\right)  .
\end{equation}

\bigskip

\section{PERTURBED COSMIC RAY CURRENT}

The linear perturbations of the components of the cosmic ray current
$\mathbf{j}_{cr1}=q_{cr}n_{cr0}\mathbf{v}_{cr1}+q_{cr}n_{cr1}\mathbf{u}_{cr}$
are equal to%
\begin{align}
4\pi j_{cr1x}  & =\frac{\omega_{pcr}^{2}}{\omega_{ccr}}E_{1y}+\frac
{\omega_{pcr}^{2}\gamma_{cr0}^{3}}{\omega_{ccr}^{2}}\frac{\partial E_{1x}%
}{\partial t}-\frac{\omega_{pcr}^{2}\gamma_{cr0}^{4}}{\omega_{ccr}^{3}}%
\frac{\partial^{2}E_{1y}}{\partial t^{2}},\\
4\pi j_{cr1y}  & =-\frac{\omega_{pcr}^{2}}{\omega_{ccr}}E_{1x}+\frac
{\omega_{pcr}^{2}\gamma_{cr0}}{\omega_{ccr}^{2}}\frac{\partial E_{1y}%
}{\partial t}+\frac{\omega_{pcr}^{2}\gamma_{cr0}^{4}}{\omega_{ccr}^{3}}%
\frac{\partial^{2}E_{1x}}{\partial t^{2}}\nonumber\\
& -4\pi q_{cr}n_{cr0}u_{cr}\frac{1}{L_{cr}}\left(  \frac{\partial}{\partial
t}\right)  ^{-1}\frac{\partial H_{cr1}}{\partial z},\nonumber\\
4\pi j_{cr1z}  & =4\pi q_{cr}n_{cr0}\frac{H_{cr1}}{L_{cr}},\nonumber
\end{align}
where $\omega_{pcr}=\left(  4\pi n_{cr0}q_{cr}^{2}/m_{cr}\right)  ^{1/2}$ is
the cosmic ray plasma frequency. When obtaining Equation (24), we have used
Equations (21) and (22) and the continuity equation for cosmic rays. The value
$H_{cr1}$ is defined by Equation (23).

\bigskip

\section{PERTURBED TOTAL CURRENT}

We find now the components of the perturbed total current $\mathbf{j}%
_{1}=\mathbf{j}_{pl1}+\mathbf{j}_{cr1}$. Adding Equations (20) and (24), we
obtain%
\begin{align}
4\pi j_{1x}  & =\left(  \frac{\omega_{pi}^{2}}{\omega_{ci}^{2}}+\frac
{\omega_{pcr}^{2}\gamma_{cr0}^{3}}{\omega_{ccr}^{2}}\right)  \frac{\partial
E_{1x}}{\partial t}-\left(  \frac{\omega_{pi}^{2}}{\omega_{ci}^{3}}%
+\frac{\omega_{pcr}^{2}\gamma_{cr0}^{4}}{\omega_{ccr}^{3}}\right)
\frac{\partial^{2}E_{1y}}{\partial t^{2}},\\
4\pi j_{1y}  & =\left(  \frac{\omega_{pi}^{2}}{\omega_{ci}^{2}}+\frac
{\omega_{pcr}^{2}\gamma_{cr0}}{\omega_{ccr}^{2}}\right)  \frac{\partial
E_{1y}}{\partial t}+\left(  \frac{\omega_{pi}^{2}}{\omega_{ci}^{3}}%
+\frac{\omega_{pcr}^{2}\gamma_{cr0}^{4}}{\omega_{ccr}^{3}}\right)
\frac{\partial^{2}E_{1x}}{\partial t^{2}}\nonumber\\
& -4\pi q_{i}n_{i0}u_{pl}\frac{1}{L}\left(  \frac{\partial}{\partial
t}\right)  ^{-1}\frac{\partial H_{i1}}{\partial z}-4\pi q_{cr}n_{cr0}%
u_{cr}\frac{1}{L_{cr}}\left(  \frac{\partial}{\partial t}\right)  ^{-1}%
\frac{\partial H_{cr1}}{\partial z},\nonumber\\
4\pi j_{1z}  & =4\pi q_{i}n_{i0}\frac{H_{i1}}{L}+4\pi q_{e}n_{e0}\frac{H_{e1}%
}{L}+4\pi q_{cr}n_{cr0}\frac{H_{cr1}}{L_{cr}},\nonumber
\end{align}
where we have used the condition of quasi-neutrality $q_{i}n_{i0}+q_{e}%
n_{e0}+q_{cr}n_{cr0}=0$.

\bigskip

\section{WAVE EQUATIONS}

To obtain the wave equations, we have to substitute the current (25) into
Maxwell's equation (10). Omitting small terms under condition $\partial
/\omega_{ci}\partial t\ll1$ and also assuming that $\gamma_{cr0}^{3}%
\ \partial/\omega_{ccr}\partial t\ll1$, we find in the one-dimensional case
\begin{equation}
c^{2}\left(  \frac{\partial}{\partial t}\right)  ^{-2}\frac{\partial^{2}E_{x}%
}{\partial z^{2}}-E_{x}=\varepsilon_{xx}E_{1x},
\end{equation}%
\begin{equation}
c^{2}\left(  \frac{\partial}{\partial t}\right)  ^{-2}\frac{\partial^{2}E_{y}%
}{\partial z^{2}}-E_{y}=\varepsilon_{yy}E_{1y}-\varepsilon_{yz}E_{1z},
\end{equation}%
\begin{equation}
0=-\varepsilon_{zy}E_{1y}+\varepsilon_{zz}E_{1z}.
\end{equation}
Here, the following notations are introduced:
\begin{align}
\varepsilon_{xx}  & \mathbf{=}\frac{\omega_{pi}^{2}}{\omega_{ci}^{2}}%
+\frac{\omega_{pcr}^{2}\gamma_{cr0}^{3}}{\omega_{ccr}^{2}},\\
\varepsilon_{yy}  & =\frac{\omega_{pi}^{2}}{\omega_{ci}^{2}}+\frac
{\omega_{pcr}^{2}\gamma_{cr0}}{\omega_{ccr}^{2}}+\left(  \omega_{pi}^{2}%
u_{pl}^{2}\frac{DG_{2}}{L}+\omega_{pcr}^{2}u_{cr}^{2}\frac{1}{L_{cr}}\right)
\left(  \frac{\partial}{\partial t}\right)  ^{-2}\frac{\partial^{2}}{\partial
z^{2}},\nonumber\\
\varepsilon_{yz}  & =\left(  \omega_{pi}^{2}u_{pl}\frac{DG_{1}}{L}%
+\omega_{pcr}^{2}u_{cr}\frac{1}{L_{cr}}\right)  \left(  \frac{\partial
}{\partial t}\right)  ^{-1}\frac{\partial}{\partial z},\nonumber\\
\varepsilon_{zy}  & =\left[  \omega_{pi}^{2}\frac{D}{L}u_{pl}\left(
G_{2}+\frac{q_{e}n_{e0}}{q_{i}n_{i0}}G_{4}\right)  +\omega_{pcr}^{2}\frac
{1}{L_{cr}}u_{cr}\right]  \left(  \frac{\partial}{\partial t}\right)
^{-1}\frac{\partial}{\partial z},\nonumber\\
\varepsilon_{zz}  & =\left(  \omega_{pe}^{2}G_{3}+\omega_{pi}^{2}G_{1}\right)
\frac{D}{L}+\omega_{pcr}^{2}\frac{1}{L_{cr}}+1.\nonumber
\end{align}
When deriving Equations (26)-(28), we have used Equations (9), (23), and (A18)
without the contribution of terms proportional to $v_{0x}$.

Equation (26) describes magnetosonic waves including the contribution of
cosmic rays at conditions under consideration. Equations (27) and (28)
describe the streaming instability due to the cosmic ray flow and thermal
instability subjected to an influence of cosmic rays. We note that when
$u_{cr}=0$, the thermal instability is defined from Equation (28),
$\varepsilon_{zz}=0$ and $E_{1z}\neq0$ (Nekrasov 2011, 2012).

\bigskip

\section{DISPERSION RELATIONS}

Equations (27) and (28) are given in their general form which permits us to
investigate different particular cases. Making use of the Fourier analysis for
perturbations proportional to $\exp\left(  ikz-i\omega t\right)  $, we obtain%
\begin{equation}
\varepsilon_{zz}\left(  \frac{k^{2}c^{2}}{\omega^{2}}-1\right)  =\varepsilon
_{yy}\varepsilon_{zz}-\varepsilon_{yz}\varepsilon_{zy}.
\end{equation}
Substituting into Equation (30) expressions for $\varepsilon_{yy}$,
$\varepsilon_{yz}$, $\varepsilon_{zy}$, and $\varepsilon_{zz}$, which are
defined by Equation (29), we find%
\begin{equation}
\varepsilon_{zz}\left(  \frac{k^{2}c^{2}}{\omega^{2}}-1-\frac{c^{2}}{c_{A}%
^{2}}\right)  \frac{\omega^{2}}{k^{2}}=\varepsilon_{zz}\alpha_{1}-\alpha
_{2}\alpha_{3},
\end{equation}
where $c_{A}$ is the Alfv\'{e}n velocity including the contribution of cosmic
rays%
\[
\frac{c^{2}}{c_{A}^{2}}=\frac{\omega_{pi}^{2}}{\omega_{ci}^{2}}+\frac
{\omega_{pcr}^{2}\gamma_{cr0}}{\omega_{ccr}^{2}}.
\]
In Equation (31), we have introduced notations
\begin{align}
\alpha_{1}  & =\omega_{pi}^{2}u_{pl}^{2}\frac{DG_{2}}{L}+\omega_{pcr}%
^{2}u_{cr}^{2}\frac{1}{L_{cr}},\\
\alpha_{2}  & =\omega_{pi}^{2}u_{pl}\frac{DG_{1}}{L}+\omega_{pcr}^{2}%
u_{cr}\frac{1}{L_{cr}},\nonumber\\
\alpha_{3}  & =\omega_{pi}^{2}\frac{D}{L}u_{pl}\left(  G_{2}+\frac{q_{e}%
n_{e0}}{q_{i}n_{i0}}G_{4}\right)  +\omega_{pcr}^{2}u_{cr}\frac{1}{L_{cr}%
}.\nonumber
\end{align}

To calculate the right-hand side of Equation (31) with values $\alpha_{1,2,3}$
given by Equation (32), it is convenient to consider the expression
$L^{2}\left(  \varepsilon_{zz}\alpha_{1}-\alpha_{2}\alpha_{3}\right)  $.
Carrying out\textbf{\ }the calculations and taking into account that
$q_{i}n_{i0}+q_{e}n_{e0}\approx0$ and $u_{cr}\gg v_{i0y}$, we obtain
\begin{align}
L^{2}\left(  \varepsilon_{zz}\alpha_{1}-\alpha_{2}\alpha_{3}\right)   &
=\omega_{pi}^{2}u_{pl}^{2}\left(  \omega_{pe}^{2}G_{2}G_{3}+\omega_{pi}%
^{2}G_{1}G_{4}\right)  D^{2}\\
& +\omega_{pcr}^{2}u_{cr}^{2}\left(  \omega_{pe}^{2}G_{3}+\omega_{pi}^{2}%
G_{1}\right)  D\frac{L}{L_{cr}}.\nonumber
\end{align}
It can be shown that the value $\omega_{pe}^{2}G_{2}G_{3}+\omega_{pi}^{2}%
G_{1}G_{4}$ acquires the simple form%
\begin{equation}
\omega_{pe}^{2}G_{2}G_{3}+\omega_{pi}^{2}G_{1}G_{4}=\omega_{pe}^{2}L,
\end{equation}
where we have taken $q_{i}=-q_{e}$ (see Equation (A19)). The value
$\omega_{pe}^{2}G_{3}+\omega_{pi}^{2}G_{1}$ can be given in the following
form:%
\begin{equation}
\omega_{pe}^{2}G_{3}+\omega_{pi}^{2}G_{1}=-\omega_{pe}^{2}D\left(  \omega
^{2}-k^{2}C_{s}^{2}\right)  ,
\end{equation}
where $C_{s}^{2}$ is given by
\begin{equation}
m_{i}DC_{s}^{2}=T_{i0}\left(  W_{i}V_{e}+2W_{i}\Omega_{ei}+V_{e}\Omega
_{ie}-V_{i}\Omega_{ei}\right)  +T_{e0}\left(  W_{e}V_{i}+2W_{e}\Omega
_{ie}+V_{i}\Omega_{ei}-V_{e}\Omega_{ie}\right)  .
\end{equation}
In the value $\varepsilon_{zz}$ on the left-hand side of Equation (31), we
neglect the contribution of the unity which has \textbf{arisen} due to the
displacement current. It is easy to see that it can be done if $\omega
_{pe}^{2}\gg\omega^{2}$ for the cold plasma and $\omega_{pe}^{2}\gg
k^{2}v_{Te}^{2}$ for the warm plasma, when the wavelength of perturbations is
much larger than the Debye length. We also omit the negligible contribution of
cosmic rays. Thus, we have
\begin{equation}
\varepsilon_{zz}=\left(  \omega_{pe}^{2}G_{3}+\omega_{pi}^{2}G_{1}\right)
\frac{D}{L}.
\end{equation}
Substituting Equations (33)-(35) and (37) into Equation (31), we derive the
following dispersion relation:
\begin{equation}
\omega^{2}\frac{c^{2}}{c_{A}^{2}}-k^{2}c^{2}=\frac{\omega_{pi}^{2}u_{pl}%
^{2}k^{2}}{\left(  \omega^{2}-k^{2}C_{s}^{2}\right)  }+\frac{\omega_{pcr}%
^{2}u_{cr}^{2}k^{2}}{\left(  \gamma_{cr0}\omega^{2}-k^{2}c_{scr}^{2}\right)
},
\end{equation}
where we also have neglected the contribution of the displacement current.
Equation (38) describes the streaming and thermal instability. Below, we
consider some particular cases.

\bigskip

\subsection{\textit{Streaming Instability without Cosmic Ray Back-Reaction }}

In this case, we neglect the contribution of the cosmic ray term in the
dispersion relation defined by Equation (38). We also set all the frequencies
$\Omega$ to zero. Then Equation (38) can be written in the form%
\begin{equation}
\omega^{2}\frac{c^{2}}{c_{Ai}^{2}}-k^{2}c^{2}=\frac{4\pi j_{ret}^{2}k^{2}%
}{\rho_{i0}\left(  \omega^{2}-k^{2}c_{s}^{2}\right)  },
\end{equation}
where $c_{s}=\left[  \gamma\left(  T_{e0}+T_{i0}\right)  /m_{i}\right]
^{1/2}$ is the plasma sound velocity, $c_{Ai}=\omega_{ci}c/\omega_{pi}$,
$\rho_{i0}=m_{i}n_{i0}$, and $j_{ret}$ is defined by Equation (16). Equation
(39) coincides with Equation (9) in the paper by Riquelme \& Spitkovsky (2010)
(see also Bell (2005)) .

\bigskip

\subsection{\textit{Streaming Instability with Cosmic Ray Back-Reaction }}

Taking into account the back-reaction of cosmic rays, Equation (38) becomes
\begin{equation}
\omega^{2}\frac{c^{2}}{c_{A}^{2}}-k^{2}c^{2}=\frac{4\pi j_{ret}^{2}k^{2}}%
{\rho_{i0}\left(  \omega^{2}-k^{2}c_{s}^{2}\right)  }+\frac{4\pi j_{cr0}%
^{2}k^{2}}{\rho_{cr0}\left(  \gamma_{cr0}\omega^{2}-k^{2}c_{scr}^{2}\right)
},
\end{equation}
where $\rho_{cr0}=$ $n_{cr0}m_{cr}$. Since $j_{cr0}=j_{ret}$, the second term
on the right-hand side of Equation (40) is considerably larger than the first
one roughly by a factor of $n_{i0}/n_{cr0}\gg1$. Thus, the back-reaction of
the streaming cosmic rays results in much more powerful instability than the
induced background plasma streaming. It should be noted that this conclusion
is satisfied for\textbf{\ }conditions under consideration. From Equation (40),
omitting the first term on the right-hand side, we can find the wave number
$k_{m}$ and growth rate $\delta_{m}$ ($\delta=-i\omega$) of the fastest
growing mode
\begin{equation}
k_{m}^{2}=\frac{8\pi j_{cr0}^{2}}{\rho_{cr0}c^{2}}\frac{\gamma_{cr0}^{-1}%
c_{A}^{2}}{\left(  \gamma_{cr0}^{-1}c_{scr}^{2}-c_{A}^{2}\right)  ^{2}%
}\left\{  \left[  1+\frac{\left(  \gamma_{cr0}^{-1}c_{scr}^{2}-c_{A}%
^{2}\right)  ^{2}}{4\gamma_{cr0}^{-1}c_{scr}^{2}c_{A}^{2}}\right]
^{1/2}-1\right\}
\end{equation}
and%
\begin{align}
\delta_{m}^{2}  & =\frac{2\pi j_{cr0}^{2}}{\rho_{cr0}c^{2}}\frac{c_{A}%
}{c_{scr}}\gamma_{cr0}^{-1/2}-\frac{4\pi j_{cr0}^{2}}{\rho_{cr0}c^{2}}%
\frac{\left(  \gamma_{cr0}^{-1}c_{scr}^{2}+c_{A}^{2}\right)  c_{A}^{2}%
\gamma_{cr0}^{-1}}{\left(  \gamma_{cr0}^{-1}c_{scr}^{2}-c_{A}^{2}\right)
^{2}}\\
& \times\left\{  \left[  1+\frac{\left(  \gamma_{cr0}^{-1}c_{scr}^{2}%
-c_{A}^{2}\right)  ^{2}}{4\gamma_{cr0}^{-1}c_{scr}^{2}c_{A}^{2}}\right]
^{1/2}-1\right\}  .\nonumber
\end{align}

Let us find asymptotical expressions for $k_{m}$ and $\delta_{m}$. In the case
$\gamma_{cr0}^{-1}c_{scr}^{2}\gg c_{A}^{2}$, we have
\begin{equation}
k_{m}^{2}=\frac{4\pi j_{cr0}^{2}}{\rho_{cr0}c^{2}}\frac{\gamma_{cr0}%
^{1/2}c_{A}}{c_{scr}^{3}},\delta_{m}^{2}=\frac{4\pi j_{cr0}^{2}}{\rho
_{cr0}c^{2}}\frac{c_{A}^{2}}{c_{scr}^{2}}.
\end{equation}
Thus, $\delta_{m}=k_{m}\left(  \gamma_{cr0}^{-1/2}c_{scr}c_{A}\right)  ^{1/2}
$. In the opposite case, $\gamma_{cr0}^{-1}c_{scr}^{2}\ll c_{A}^{2}$, we
obtain%
\begin{equation}
k_{m}^{2}=\frac{4\pi j_{cr0}^{2}}{\rho_{cr0}c^{2}}\frac{\gamma_{cr0}^{-1/2}%
}{c_{scr}c_{A}},\delta_{m}^{2}=\frac{4\pi j_{cr0}^{2}}{\rho_{cr0}c^{2}}%
\gamma_{cr0}^{-1}.
\end{equation}
The relation between $\delta_{m}$ and $k_{m}$ is the same as for solutions
(43). From Equations (43) and (44), we can write expressions for $k_{m}^{2}$
and $\delta_{m}^{2}$, which unite both limiting cases
\begin{equation}
k_{m}^{2}=\frac{4\pi j_{cr0}^{2}}{\rho_{cr0}c^{2}}\frac{\gamma_{cr0}%
^{-1/2}c_{A}}{c_{scr}\left(  \gamma_{cr0}^{-1}c_{scr}^{2}+c_{A}^{2}\right)
},\delta_{m}^{2}=\frac{4\pi j_{cr0}^{2}}{\rho_{cr0}c^{2}}\frac{\gamma
_{cr0}^{-1}c_{A}^{2}}{\gamma_{cr0}^{-1}c_{scr}^{2}+c_{A}^{2}}.
\end{equation}
In the resonance case, $\gamma_{cr0}^{-1}c_{scr}^{2}\approx c_{A}^{2}$, we
find from Equations (41) and (42)%
\begin{equation}
k_{m}^{2}=\frac{\pi j_{cr0}^{2}}{\rho_{cr0}c^{2}}\frac{1}{c_{scr}^{2}}%
,\delta_{m}^{2}=\frac{\pi j_{cr0}^{2}}{\rho_{cr0}c^{2}}\gamma_{cr0}^{-1}.
\end{equation}
As we see, magnitudes given by\textbf{\ }Equation (45) in the resonance case
are only twice as large as those in Equation (46). Thus, Equation (45) can be
applied to a good accuracy for any relation between $\gamma_{cr0}^{-1}%
c_{scr}^{2}$ and $c_{A}^{2}$.

\bigskip

\subsection{\textit{Thermal Instability with Cosmic Ray Back-reaction}}

From Equation (38), it is clear that the thermal instability is described by
equation%
\begin{equation}
\omega^{2}-k^{2}C_{s}^{2}=0,
\end{equation}
where $C_{s}^{2}$ is given by Equation (36). If we set $u_{cr}=0$, then
Equation (47) is satisfied. In the case $u_{cr}\neq0$, the value $\left(
\omega^{2}-k^{2}C_{s}^{2}\right)  ^{-1}$ is multiplied by a small coefficient
in Equation (38) in comparison with the second term on the right-hand side of
this equation. Therefore, Equation (47) is kept. Thus, cosmic rays do not
influence on the thermal instability under conditions considered in this paper.

We set in Equation (47) $T_{i0}=T_{e0}=T_{0}$. Then, this equation coincides
with Equation (47) in the paper by Nekrasov (2011) without the inertia term.
We note that in the last paper the perturbation of the thermal energy exchange
frequency has been taken into account. We further take $\Omega_{ie}%
=\Omega_{ei}=\Omega_{\epsilon}$. Then, we have
\begin{equation}
\delta^{2}\left(  \delta^{2}+\beta_{3}\delta+\beta_{4}\right)  +\frac
{1}{2\gamma}k^{2}c_{s}^{2}\left(  2\gamma\delta^{2}+\beta_{1}\delta+\beta
_{2}\right)  =0,
\end{equation}
where, as above, $\delta=-i\omega$ and $c_{s}=\left(  2\gamma T_{0}%
/m_{i}\right)  ^{1/2}$. The following notations are introduced:
\begin{align}
\beta_{1}  & =\left(  \gamma+1\right)  \left(  \Omega_{\chi}+\Omega
_{Te}+\Omega_{Ti}\right)  -\Omega_{ne}-\Omega_{ni}+4\gamma\Omega_{\epsilon},\\
\beta_{2}  & =\left(  \Omega_{\chi}+\Omega_{Te}-\Omega_{ne}\right)
\Omega_{Ti}+\left(  \Omega_{Ti}-\Omega_{ni}\right)  \left(  \Omega_{\chi
}+\Omega_{Te}\right) \nonumber\\
& +2\left(  \Omega_{\chi}+\Omega_{Te}-\Omega_{ne}+\Omega_{Ti}-\Omega
_{ni}\right)  \Omega_{\epsilon},\nonumber\\
\beta_{3}  & =\Omega_{\chi}+\Omega_{Te}+\Omega_{Ti}+2\Omega_{\epsilon
},\nonumber\\
\beta_{4}  & =\left(  \Omega_{\chi}+\Omega_{Te}\right)  \Omega_{Ti}+\left(
\Omega_{\chi}+\Omega_{Te}+\Omega_{Ti}\right)  \Omega_{\epsilon}.\nonumber
\end{align}
The frequencies $\Omega_{Te,i}$, $\Omega_{ne,i}$, $\Omega_{ei,ie}$, and
$\Omega_{\chi}$ are given by Equation (A7). In the general form, Equation (48)
can be solved numerically. We note that $\Omega_{\chi}=\left(  \gamma
-1\right)  \chi_{e0}k^{2}/n_{e0}$.

We now treat Equations (48) and (49) in the limit $\Omega_{\epsilon}\gg
\Omega_{\chi},\Omega_{Te,i},\Omega_{ne,i}$. In the short wavelength limit,
$k^{2}c_{s}^{2}\gg\delta^{2}$, the dispersion relation has the form%
\begin{equation}
\delta^{2}+2\Omega_{\epsilon}\delta+\frac{1}{\gamma}\Omega_{T,n}%
\Omega_{\epsilon}=0,
\end{equation}
where $\Omega_{T,n}=\left(  \Omega_{\chi}+\Omega_{Te}-\Omega_{ne}+\Omega
_{Ti}-\Omega_{ni}\right)  $. Solution of Equation (50) is the following:%
\begin{equation}
\delta=-\frac{1}{2\gamma}\Omega_{T,n}.
\end{equation}
This solution corresponds to Field's isobaric solution (Field 1965). In the
long wavelength limit, $k^{2}c_{s}^{2}\ll\delta^{2}$, we have equation
\begin{equation}
\delta^{2}+2\Omega_{\epsilon}\delta+\Omega_{T}\Omega_{\epsilon}=0,
\end{equation}
where $\Omega_{T}=\left(  \Omega_{\chi}+\Omega_{Te}+\Omega_{Ti}\right)  $.
Solution of Equation (52) is%
\begin{equation}
\delta=-\frac{1}{2}\Omega_{T},
\end{equation}
which corresponds to Parker's isochoric solution (Parker 1953).

\bigskip

\subsection{\textit{Thermal Instability without Cosmic Ray Back-reaction}}

For ultrarelativistic cosmic rays, $\gamma_{cr0}\rightarrow\infty$, their
back-reaction is absent (see Appendix B). In this case, Equation (38) takes
the form%

\begin{equation}
\delta^{2}+k^{2}C_{s}^{2}=\omega_{pi}^{2}\frac{n_{cr0}^{2}}{n_{i0}^{2}}%
\frac{k^{2}c_{Ai}^{2}}{\left(  \delta^{2}+k^{2}c_{Ai}^{2}\right)  },
\end{equation}
where%
\begin{equation}
C_{s}^{2}=\frac{1}{2\gamma}c_{s}^{2}\frac{2\gamma\delta^{2}+\beta_{1}%
\delta+\beta_{2}}{\delta^{2}+\beta_{3}\delta+\beta_{4}}.
\end{equation}
In Equation (54), we have assumed that $q_{cr}=q_{i}$ and $u_{cr}\approx c$.
We also consider that $n_{i0}=n_{e0}$. In the low-frequency regime,
$\delta^{2}\ll k^{2}c_{Ai}^{2}$, Equation (54) together with Equation (55) is
given by%
\begin{equation}
2\gamma\delta^{2}+k^{2}c_{s}^{2}\frac{2\gamma\delta^{2}+\beta_{1}\delta
+\beta_{2}}{\delta^{2}+\beta_{3}\delta+\beta_{4}}=2\gamma\omega_{pi}^{2}%
\frac{n_{cr0}^{2}}{n_{i0}^{2}}.
\end{equation}
If we assume that $\delta^{2}\gg\omega_{pi}^{2}n_{cr0}^{2}/n_{i0}^{2}$, then
we return to the case considered in the previous section. In this case, the
plasma return current plays no the role. When the opposite condition,
$\delta^{2}\ll\omega_{pi}^{2}n_{cr0}^{2}/n_{i0}^{2}$, is satisfied then
Equation (56) takes the form%
\begin{equation}
\frac{2\gamma\delta^{2}+\beta_{1}\delta+\beta_{2}}{\delta^{2}+\beta_{3}%
\delta+\beta_{4}}=2\gamma\frac{\omega_{pi}^{2}}{k^{2}c_{s}^{2}}\frac
{n_{cr0}^{2}}{n_{i0}^{2}}=a.
\end{equation}
We see from Equation (57) that in the limiting case $a\ll1$ ($a\gg1$) the
nominator (denominator) tends to zero. In the case $a\sim1$, the dispersion
relation is modified, however the qualitative character of the thermal
instability does not change. Thus, the plasma return current does not
influence on the thermal instability in the low-frequency regime. In the
high-frequency regime, $\delta^{2}\gg k^{2}c_{Ai}^{2}$, Equation (54) is the
following:%
\begin{equation}
\frac{2\gamma\delta^{2}+\beta_{1}\delta+\beta_{2}}{\delta^{2}+\beta_{3}%
\delta+\beta_{4}}=2\gamma\frac{\omega_{pi}^{2}}{\delta^{2}}\frac{n_{cr0}^{2}%
}{n_{i0}^{2}}\frac{c_{Ai}^{2}}{c_{s}^{2}}=b,
\end{equation}
where we have assumed that $\delta^{2}\ll\omega_{pi}kc_{Ai}n_{cr0}/n_{i0}$.
Again, if $b\ll1$ ($b\gg1$), then the nominator (denominator) on the left-hand
side of this equation tends to zero. Thus, the plasma return current has no
effect on the thermal instability in these limiting cases. When $b\sim1$, we
have solution
\begin{equation}
\delta\sim\omega_{pi}\frac{n_{cr0}}{n_{i0}}\frac{c_{Ai}}{c_{s}}.
\end{equation}
The left-hand side of Equation (58) is of the order of the unity and does not
describe the thermal instability for the solution given by Equation (59).

\bigskip

\section{DISCUSSION}

In this paper, no conditions have been used for the background plasma except
for $\omega_{cj}^{2}\gg\partial^{2}/\partial t^{2}$, which is usually
satisfied in astrophysical settings. For cosmic rays, we have assumed that
$\gamma_{cr0}^{3}\ \partial/\omega_{ccr}\partial t\ll1$ (see Section 9). This
condition can be satisfied for moderately relativistic cosmic rays. However,
it can be violated for ultrarelativistic cosmic rays. Using the growth rate
(45) in the case for example $\gamma_{cr0}^{-1}c_{scr}^{2}\gtrsim c_{A}^{2}$,
the last condition can be written in the form%
\begin{equation}
\gamma_{cr0}^{3}\ \left(  \frac{n_{cr}}{n_{i0}}\right)  ^{1/2}\frac{u_{cr}%
}{c_{scr}}\ll1,
\end{equation}
where we have assumed that $c_{A}\sim c_{Ai}$ and $\omega_{ccr}\sim\omega
_{ci}$. In the limit $u_{cr}\rightarrow c$, Equation (60), taking into account
that $T_{cr}\ll m_{cr}c^{2}$, can be violated for sufficiently dense cosmic
rays. In the opposite case, $\gamma_{cr0}^{-1}c_{scr}^{2}\ll c_{A}^{2}$, the
corresponding condition is given by%
\[
\gamma_{cr0}^{5/2}\ \left(  \frac{n_{cr}}{n_{i0}}\right)  ^{1/2}\frac{u_{cr}%
}{c_{Ai}}\ll1.
\]

We see from Equation (38) that at conditions under consideration and for
one-dimensional perturbations along the background magnetic field cosmic rays
do not influence on the thermal instability (see Sections 10.3 and 10.4). At
the same time, the back-reaction of cosmic rays results in the aperiodic
streaming instability much more powerful than that due to the return current
of the background plasma. The maximal growth rate is achieved for sufficiently
cold cosmic rays and large magnetic fields such that $c_{A}^{2}\gtrsim
\gamma_{cr0}^{-1}c_{scr}^{2}$. In this case, the growth rate is equal to%
\[
\delta_{\max}\approx\omega_{pcr}\frac{u_{cr}}{c}\gamma_{cr0}^{-1/2}
\]
and the wave number is
\[
k_{m}=\omega_{pcr}\frac{u_{cr}}{c}\frac{\gamma_{cr0}^{-1/4}}{\left(
c_{scr}c_{A}\right)  ^{1/2}}.
\]
Thus in particular cases such as considered here as an example, the cosmic ray
back-reaction must be certainly invoked to study cosmic ray streaming instabilities.

We see that the thermal instability (Equation (47)) is not influenced by the
action of cosmic rays in the\textbf{\ }model under consideration, if we take
into account the cosmic ray back-reaction. The multi-fluid dispersion relation
includes the ion inertia and has a general form except for $T_{i0}=T_{e0}$. In
the limit of fast thermal energy exchange in which $\Omega_{\epsilon}$ is much
larger than all other frequencies, the isobaric and isochoric growth rates
have been obtained (Equations (51) and (53)). Ultrarelativistic cosmic rays do
not experience back-reaction. In this case, the plasma term with the return
current is contained in the dispersion relation (Equation (54)). However, as
we have shown, this term also does not influence on the thermal instability.

We note that all instabilities considered in this paper are connected with the
particle dynamics along the background magnetic field.

We have explored the situation in which cosmic rays drift across the
background magnetic field. This model has been considered by Riquelme \&
Spitkovsky (2010) for the problem of the magnetic field amplification in the
upstream region of the supernova remnant shocks. However, such a model can
also be applied to the ICM where cosmic rays are the important ingredient
(Loewenstein et al. 1991; Guo \& Oh 2008; Sharma et al. 2009, 2010). In
another model, cosmic rays drift along the magnetic field. This case has been
investigated by Bell (2004) (see also Riquelme \& Spitkovsky 2009). In both
cases the growth rates are the same (Bell 2004, 2005; Riquelme \& Spitkovsky
2010). Such a situation can also be encountered in the ICM. In the papers by
Bell (2004, 2005) and Riquelme \& Spitkovsky (2010), the return current of the
background plasma has only been involved in the analytical treatments. The
cosmic ray back-reaction has been included in the numerical analysis and found
to determine the saturation of the instability (Riquelme \& Spitkovsky 2009,
2010). However, the influence of the cosmic ray back-reaction on the growth
rate remained unknown.

As we have obtained in this paper, the cosmic ray back-reaction drives the
instability whose growth rate is proportional to $n_{cr0}^{-1/2}$, but not
$n_{i0}^{-1/2}$ as the one\textbf{\ }due to the return plasma current.
Therefore, this instability can produce much larger magnetic field
amplification in both the upstream medium of shocks and ICM. In unstable
regions, an enhanced X-ray radiation must be observed.

\bigskip

\section{ASTROPHYSICAL\ IMPLICATIONS}

Our linear analysis of the instabilities related to the current-driven
instability by cosmic rays is applicable to a variety of environments.
Although such a type of instability was suggested originally for the magnetic
field amplification in the shocks by the supernova, we think, wherever there
is a strong cosmic ray streaming, this instability may play a significant
role. For example, if the supernova driven shock propagates through a hot and
low density medium (i.e., superbubbles), then the current-driven instability
may exist. Even at larger scales, such as shocks in the ICM, we may expect
this instability under some conditions. Most of the previous analytical
studies are restricted to cosmic rays drifting along the ambient magnetic
field without their back-reaction and possible thermal effects. Interestingly,
our analysis shows that inclusion of back-reaction will lead to a much
stronger instability in comparison to the previous studies where this effect
is\textbf{\ }neglected. So, we expect that the magnetic field is amplified to
a larger value in the presence of the back-reaction of cosmic rays. This
implies more confinement of cosmic rays with excited turbulent motions in the
non-linear regime and accordingly the acceleration of cosmic rays to higher energies.

In some of supernova remnants such as IC 443, SN 1006, Kepler, Tycho and etc.,
the driven shocks are propagating in their partially ionized ambient medium.
This was a good motivation to extend CRCD instability from MHD approach to a
two-fluid case, by considering ions and neutrals as two separate fluids where
they can exchange momentum via collisions (e.g., Reville et al. 2007; see also
Bykov \& Toptygin 2005). It was shown the instability is getting slower rate
because of collisions of ions with neutrals, in particular when the cosmic ray
flux is not very strong. However, the back-reaction of cosmic rays has not
been considered\textbf{\ }by\textbf{\ }Reville et al. (2007). Considering the
finding that the growth rate is significantly enhanced in the presence of
cosmic ray back-reaction in a two-fluid system consisting of the ions and
electrons,\ one may naturally expect such an effect in a three-fluid system
consisting of the ions, electrons, and neutrals. It deserves a further study,
but we may expect that the stabilizing effect of the ion-neutral collisions
will be compensated by the back-reaction of cosmic rays.

\bigskip

\section{CONCLUSION}

Using the multi-fluid approach, we have investigated streaming and thermal
instabilities of the electron-ion plasma with homogeneous cold cosmic rays
drifting across the background magnetic field. We have taken into account the
return current of the background plasma and the back-reaction of cosmic rays
for one-dimensional perturbations along the magnetic field. It has been shown
that the cosmic ray back-reaction results in a streaming instability having
considerably larger growth rate than that due to usually treated return
current of the background plasma. The maximal growth rates and corresponding
wave numbers have been found.

The thermal instability has been shown to be not subjected to the action of
cosmic rays in the model under consideration. The dispersion relation for the
thermal instability in the multi-fluid approach has been derived which
includes the ion inertia. In the limit of fast thermal energy exchange between
electrons and ions the isobaric and isochoric growth rates have been obtained.

The results of this paper can be useful for the investigation of the
electron-ion astrophysical objects such as galaxy clusters including the
dynamics of streaming cosmic rays.

\bigskip

\section{\bigskip REFERENCES}

Alfv\'{e}n, H. 1939, Phys. Rev., 55, 425

Begelman, M. C., \& Zweibel, E. G. 1994, ApJ, 431, 689

Bell, A. R. 2004, MNRAS, 353, 550

Bell, A. R. 2005, MNRAS, 358, 181

Bogdanovi\'{c}, T., Reynolds, C. S., Balbus, S. A., \& Parrish, I. J. 2009,
ApJ, 704, 211

Braginskii, S. I. 1965, Rev. Plasma Phys., 1, 205

Bykov, A. M., \& Toptygin, I. N. 2005, Astronomy Letters, 31, 839

Cavagnolo, K. W., Donahue, M., Voil, G. M., \& Sun, M. 2008, ApJ, 683, L107

Conselice, C. J., Gallagher, J. S., III, Wyse R. F. G. 2001, AJ, 122, 2281

Dzhavakhishvili, D. I., \& Tsintsadze, N. L. 1973, Sov. Phys. JEPT, 37, 666;
1973, Zh. Eksp. Teor. Fiz. 64, 1314

Everett, J. E., Zweibel, E. G., Benjamin, R. A., McCammon, D., Rocks, L., \&
Gallagher, J. S. 2008, ApJ, 674, 258

Ferland, G. J., Fabian, A. C., Hatch, N. A., Johnstone, R. M., Porter, R. L.,
van Hoof, P. A. M., \& Williams, R. J. R. 2009, MNRAS, 392, 1475

Field, G.B. 1965, ApJ, 142, 531

Field, G. B., Goldsmith, D. W., \& Habing, H. J. 1969, ApJ, 155, L149

Gammie, C. F. 1996, ApJ, 457, 355

Goldsmith, D. W., Habing, H. J., \& Field, G. B. 1969, ApJ, 158, 173

Guo, F., \& Oh, S. P. 2008, MNRAS, 384, 251

Khajenabi, F. 2012, Astrophys. Space Sci., 337, 247

Kulsrud, R., \& Pearce, W. P. 1969, ApJ, 156, 445

Kuwabara, T., \& Ko, C.-M. 2006, ApJ, 636, 290

Loewenstein, M., Zweibel, E. G., \& Begelman, M. C. 1991, ApJ, 377, 392

Lontano, M., Bulanov, S., \& Koga, J. 2002, AIP Conf. Proc., 611, 157

Nekrasov, A. K. 2008, Phys. Plasmas, 15, 032907

Nekrasov, A. K. 2009a, MNRAS, 400, 1574

Nekrasov, A. K. 2009b, ApJ, 695, 46

Nekrasov, A. K. 2009c, ApJ, 704, 80

Nekrasov, A. K. 2011, ApJ, 739, 88

Nekrasov, A. K. 2012, MNRAS, 419, 522

Nekrasov, A. K., \& Shadmehri, M. 2010, ApJ, 724,\textbf{\ }1165

Nekrasov, A. K., \& Shadmehri, M. 2011, Ap\&SS, 333, 477

O'Dea, C. P. et al. 2008, ApJ, 681, 1035

Parker, E. N. 1953, ApJ, 117, 431

Parker, E. N. 1966, ApJ, 145, 811

Parrish, I. J., Quataert, E., \& Sharma, P. 2009, ApJ, 703, 96

Reville, B., \& Bell, A. R. 2012, MNRAS, 419, 2433

Reville, B., Kirk, J. G., Duffy, P., \& O'Sullivan, S. 2007, A\&A, 475, 435

Riquelme, M. A., \& Spitkovsky, A. 2009, ApJ, 694, 626

Riquelme, M. A., \& Spitkovsky, A. 2010, ApJ, 717, 1054

Salom\'{e} P. et al., 2006, A\&A, 454, 437

Shadmehri, M. 2009, MNRAS, 397, 1521

Sharma, P., Chandran, B. D. G., Quataert, E., \& Parrish, I. J. 2009, ApJ,
699, 348

Sharma, P., Parrish, I. J., \& Quataert, E. 2010, ApJ, 720, 652

Snodin, A. P., Brandenburg, A., Mee, A. J., \& Shukurov, A. 2006, MNRAS, 373, 643

Toepfer, A. J. 1971, Phys. Rev. A, 3, 1444

Tozzi, P., \& Norman, C. 2001, ApJ, 546, 63

Wagner, A. Y., Falle, S. A. E. G., Hartquist, T. W., \& Pittard, J. M. 2005,
A\& A, 430, 567

Yusef-Zadeh, F., Wardle, M., \& Roy, S. 2007, ApJ, 665, L123

Zweibel, E. G. 2003, ApJ, 587, 625

Zweibel, E. G., \& Everett, J. E. 2010, ApJ, 709, 1412

\bigskip
\begin{appendix}

\section{APPENDIX}

\bigskip

\subsection{\textit{Perturbed Velocities of Ions and Electrons }}

We put in Equation (1) $\mathbf{v}_{j}=\mathbf{v}_{j0}+\mathbf{v}_{j1}$,
$p_{j}=p_{j0}+p_{j1}$, $\mathbf{E=E}_{0}+\mathbf{E}_{1}$, $\mathbf{B=B}%
_{0}+\mathbf{B}_{1}$. For perturbations depending only on the $z$-coordinate,
we have $\mathbf{v}_{j0}\cdot\mathbf{\nabla=}0$. Then the linearized Equation
(1) takes the form
\begin{equation}
\frac{\partial\mathbf{v}_{j1}}{\partial t}=-\frac{\mathbf{\nabla}T_{j1}}%
{m_{j}}-\frac{T_{j0}}{m_{j}}\frac{\mathbf{\nabla}n_{j1}}{n_{j0}}%
+\mathbf{F}_{j1}\mathbf{+}\frac{q_{j}}{m_{j}c}\mathbf{v}_{j1}\times
\mathbf{B}_{0},\tag{A1}%
\end{equation}
where we have used that $p_{j1}=n_{j0}T_{j1}+n_{j1}T_{j0}$ ($n_{j}%
=n_{j0}+n_{j1}$, $T_{j}=T_{j0}+T_{j1}$) and introduced the notation%
\begin{equation}
\mathbf{F}_{j1}=\frac{q_{j}}{m_{j}}\mathbf{E}_{1}\mathbf{+}\frac{q_{j}}%
{m_{j}c}\mathbf{v}_{j0}\times\mathbf{B}_{1}.\tag{A2}%
\end{equation}
We find from Equation (A1) the following equations for $v_{j1x,y}$:%
\begin{align}
\left(  \frac{\partial^{2}}{\partial t^{2}}+\omega_{cj}^{2}\right)  v_{j1x}  &
=\omega_{cj}F_{j1y}+\frac{\partial F_{j1x}}{\partial t},\tag{A3}\\
\left(  \frac{\partial^{2}}{\partial t^{2}}+\omega_{cj}^{2}\right)  v_{j1y}  &
=-\omega_{cj}F_{j1x}+\frac{\partial F_{j1y}}{\partial t}.\nonumber
\end{align}
Applying $\partial/\partial t$ to the $z$-component of Equation (A1) and using
the linearized continuity equation (2), we obtain%
\begin{equation}
\left(  \frac{\partial^{2}}{\partial t^{2}}-\frac{T_{j0}}{m_{j}}\frac
{\partial^{2}}{\partial z^{2}}\right)  v_{j1z}=-\frac{1}{m_{j}}\frac
{\partial^{2}T_{j1}}{\partial z\partial t}+\frac{\partial F_{j1z}}{\partial
t}.\tag{A4}%
\end{equation}

\bigskip

\subsection{\textit{Perturbed Temperatures of Ions and Electrons }}

Let us find now equations for the temperature perturbations. Linearized
versions of Equations (3) and (4) for one-dimensional perturbations are given
by
\begin{align}
D_{1i}T_{i1}  & =-C_{1i}\frac{n_{i1}}{n_{i0}}+\Omega_{ie}T_{e1},\tag{A5}\\
D_{1e}T_{e1}  & =-C_{1e}\frac{n_{e1}}{n_{e0}}+\Omega_{ei}T_{i1},\nonumber
\end{align}
where notations are introduced%
\begin{align}
D_{1i}  & =\frac{\partial}{\partial t}+\Omega_{Ti}+\Omega_{ie},C_{1i}%
=T_{i0}\left[  -\left(  \gamma-1\right)  \frac{\partial}{\partial t}%
+\Omega_{ni}\right]  ,\tag{A6}\\
D_{1e}  & =\frac{\partial}{\partial t}+\Omega_{\chi}+\Omega_{Te}+\Omega
_{ei},C_{1e}=T_{e0}\left[  -\left(  \gamma-1\right)  \frac{\partial}{\partial
t}+\Omega_{ne}\right]  .\nonumber
\end{align}
The frequencies $\Omega$ in Equations (A5) and (A6) are the following:%
\begin{align}
\Omega_{Ti}  & =\left(  \gamma-1\right)  \frac{\partial%
\mathcal{L}%
_{i}\left(  n_{i0},T_{i0}\right)  }{n_{i0}\partial T_{i0}},\Omega_{ni}=\left(
\gamma-1\right)  \frac{\partial%
\mathcal{L}%
_{i}\left(  n_{i0},T_{i0}\right)  }{T_{i0}\partial n_{i0}},\tag{A7}\\
\Omega_{Te}  & =\left(  \gamma-1\right)  \frac{\partial%
\mathcal{L}%
_{e}\left(  n_{e0},T_{e0}\right)  }{n_{e0}\partial T_{e0}},\Omega_{ne}=\left(
\gamma-1\right)  \frac{\partial%
\mathcal{L}%
_{e}\left(  n_{e0},T_{e0}\right)  }{T_{e0}\partial n_{e0}},\Omega_{\chi
}=-\left(  \gamma-1\right)  \frac{\chi_{e0}}{n_{e0}}\frac{\partial^{2}%
}{\partial z^{2}},\nonumber\\
\Omega_{ie}  & =\nu_{ie}^{\varepsilon}\left(  n_{e0},T_{e0}\right)
,\Omega_{ei}=\nu_{ei}^{\varepsilon}\left(  n_{i0},T_{e0}\right)  .\nonumber
\end{align}
When deriving Equation (A5), we have used Equation (17) and Equations (2) and
(5) in their linearized form. From Equation (A5), we can express temperature
perturbations through the number density perturbations%
\begin{align}
DT_{i1}  & =-D_{1e}C_{1i}\frac{n_{i1}}{n_{i0}}-\Omega_{ie}C_{1e}\frac{n_{e1}%
}{n_{e0}},\tag{A8}\\
DT_{e1}  & =-D_{1i}C_{1e}\frac{n_{e1}}{n_{e0}}-\Omega_{ei}C_{1i}\frac{n_{i1}%
}{n_{i0}},\nonumber
\end{align}
where
\begin{equation}
D=D_{1i}D_{1e}-\Omega_{ie}\Omega_{ei}.\tag{A9}%
\end{equation}
To proceed further, we apply operator $\partial/\partial t$ to Equation (A8)
and use the continuity equation. As a result, we obtain
\begin{align}
D\frac{\partial T_{i1}}{\partial t}  & =D_{1e}C_{1i}\frac{\partial v_{i1z}%
}{\partial z}+\Omega_{ie}C_{1e}\frac{\partial v_{e1z}}{\partial z},\tag{A10}\\
D\frac{\partial T_{e1}}{\partial t}  & =D_{1i}C_{1e}\frac{\partial v_{e1z}%
}{\partial z}+\Omega_{ei}C_{1i}\frac{\partial v_{i1z}}{\partial z}.\nonumber
\end{align}
These equations, we have to introduce into Equation (A4).

\bigskip

\subsection{\textit{Equations for longitudinal Velocities }$v_{i1z}$ and
$v_{e1z}$}

Let us rewrite Equation (A4) for each component of species and use Equation
(A10). Then we obtain
\begin{align}
L_{1i}v_{i1z}+L_{2i}v_{e1z}  & =D\frac{\partial F_{i1z}}{\partial t}%
,\tag{A11}\\
L_{1e}v_{e1z}+L_{2e}v_{i1z}  & =D\frac{\partial F_{e1z}}{\partial t}.\nonumber
\end{align}
Here, the following notations are introduced:%
\begin{align}
L_{1i}  & =D\frac{\partial^{2}}{\partial t^{2}}+\frac{1}{m_{i}}\left(
D_{1e}C_{1i}-T_{i0}D\right)  \frac{\partial^{2}}{\partial z^{2}},L_{2i}%
=\frac{1}{m_{i}}\Omega_{ie}C_{1e}\frac{\partial^{2}}{\partial z^{2}}%
,\tag{A12}\\
L_{1e}  & =D\frac{\partial^{2}}{\partial t^{2}}+\frac{1}{m_{e}}\left(
D_{1i}C_{1e}-T_{e0}D\right)  \frac{\partial^{2}}{\partial z^{2}},L_{2e}%
=\frac{1}{m_{e}}\Omega_{ei}C_{1i}\frac{\partial^{2}}{\partial z^{2}}.\nonumber
\end{align}
From Equation (A11), we find equations for $v_{i1z}$ and $v_{e1z}$%
\begin{equation}
Lv_{i1z}=H_{i1},Lv_{e1z}=H_{e1},\tag{A13}%
\end{equation}
where%
\begin{align}
L  & =\left(  L_{1i}L_{1e}-L_{2i}L_{2e}\right)  ,\tag{A14}\\
H_{i1}  & =D\frac{\partial}{\partial t}\left(  L_{1e}F_{i1z}-L_{2i}%
F_{e1z}\right)  ,\nonumber\\
H_{e1}  & =D\frac{\partial}{\partial t}\left(  L_{1i}F_{e1z}-L_{2e}%
F_{i1z}\right)  .\nonumber
\end{align}

\bigskip

\subsection{\textit{Simplification of Operators defining }$v_{i,e1z}$}

Let us introduce notations%
\begin{align}
W_{i}  & =\gamma\frac{\partial}{\partial t}+\Omega_{Ti}-\Omega_{ni}%
,V_{i}=\frac{\partial}{\partial t}+\Omega_{Ti},\tag{A15}\\
W_{e}  & =\gamma\frac{\partial}{\partial t}+\Omega_{\chi}+\Omega_{Te}%
-\Omega_{ne},V_{e}=\frac{\partial}{\partial t}+\Omega_{\chi}+\Omega
_{Te}.\nonumber
\end{align}
Then the following operators take the form%
\begin{align}
D  & =V_{i}V_{e}+\Omega_{ei}V_{i}+\Omega_{ie}V_{e},\tag{A16}\\
-\frac{1}{T_{i0}}\left(  D_{1e}C_{1i}-T_{i0}D\right)   & =W_{i}\left(
V_{e}+\Omega_{ei}\right)  +\Omega_{ie}V_{e},\nonumber\\
-\frac{1}{T_{e0}}\left(  D_{1i}C_{1e}-T_{e0}D\right)   & =W_{e}\left(
V_{i}+\Omega_{ie}\right)  +\Omega_{ei}V_{i},\nonumber\\
L  & =D^{2}\frac{\partial^{4}}{\partial t^{4}}+L_{1}D\frac{\partial^{4}%
}{\partial z^{2}\partial t^{2}}+\frac{T_{i0}T_{e0}}{m_{i}m_{e}}L_{2}%
\frac{\partial^{4}}{\partial z^{4}},\nonumber
\end{align}
where%
\begin{align}
L_{1}  & =-\frac{T_{e0}}{m_{e}}\left[  W_{e}\left(  V_{i}+\Omega_{ie}\right)
+V_{i}\Omega_{ei}\right]  -\frac{T_{i0}}{m_{i}}\left[  W_{i}\left(
V_{e}+\Omega_{ei}\right)  +V_{e}\Omega_{ie}\right]  ,\tag{A17}\\
L_{2}  & =W_{i}W_{e}V_{i}V_{e}+W_{i}V_{i}\left(  W_{e}+V_{e}\right)
\Omega_{ei}+W_{e}V_{e}\left(  W_{i}+V_{i}\right)  \Omega_{ie}\nonumber\\
& +W_{i}V_{i}\Omega_{ei}^{2}+W_{e}V_{e}\Omega_{ie}^{2}+\left(  W_{i}%
V_{e}+W_{e}V_{i}\right)  \Omega_{ie}\Omega_{ei}.\nonumber
\end{align}

We further have%
\begin{align}
\left(  D\frac{\partial}{\partial t}\right)  ^{-1}H_{i1}  & =\frac{q_{i}%
}{m_{i}}G_{1}\left(  E_{1z}\mathbf{+}\frac{v_{0x}}{c}B_{1y}\right)
-\frac{v_{i0y}}{c}\frac{q_{i}}{m_{i}}G_{2}B_{1x},\tag{A18}\\
\left(  D\frac{\partial}{\partial t}\right)  ^{-1}H_{e1}  & =\frac{q_{e}%
}{m_{e}}G_{3}\left(  E_{1z}\mathbf{+}\frac{v_{0x}}{c}B_{1y}\right)
-\frac{v_{i0y}}{c}\frac{q_{i}}{m_{i}}G_{4}B_{1x},\nonumber
\end{align}
where notations are introduced%
\begin{align}
G_{1}  & =D\frac{\partial^{2}}{\partial t^{2}}-\frac{T_{e0}}{m_{e}}\left[
W_{e}V_{i}+W_{e}\Omega_{ie}+V_{i}\Omega_{ei}-\frac{q_{e}}{q_{i}}\left(
W_{e}-V_{e}\right)  \Omega_{ie}\right]  \frac{\partial^{2}}{\partial z^{2}%
},\tag{A19}\\
G_{2}  & =D\frac{\partial^{2}}{\partial t^{2}}-\frac{T_{e0}}{m_{e}}\left(
W_{e}V_{i}+W_{e}\Omega_{ie}+V_{i}\Omega_{ei}\right)  \frac{\partial^{2}%
}{\partial z^{2}},\nonumber\\
G_{3}  & =D\frac{\partial^{2}}{\partial t^{2}}-\frac{T_{i0}}{m_{i}}\left[
W_{i}V_{e}+W_{i}\Omega_{ei}+V_{e}\Omega_{ie}-\frac{q_{i}}{q_{e}}\left(
W_{i}-V_{i}\right)  \Omega_{ei}\right]  \frac{\partial^{2}}{\partial z^{2}%
},\nonumber\\
G_{4}  & =\frac{T_{i0}}{m_{e}}\left(  W_{i}-V_{i}\right)  \Omega_{ei}%
\frac{\partial^{2}}{\partial z^{2}}.\nonumber
\end{align}
In Equation (A18), we have used expressions (see Equation (A2))
\begin{align}
F_{i1z}  & =\frac{q_{i}}{m_{i}}\left(  E_{1z}\mathbf{+}\frac{v_{0x}}{c}%
B_{1y}-\frac{v_{i0y}}{c}B_{1x}\right)  ,\tag{A20}\\
F_{e1z}  & =\frac{q_{e}}{m_{e}}\left(  E_{1z}\mathbf{+}\frac{v_{0x}}{c}%
B_{1y}\right)  ,\nonumber
\end{align}
where $v_{0x}=cE_{0y}/B_{0}$.

\bigskip

\section{\bigskip APPENDIX}

\subsection{\textit{Perturbed Velocity of Cosmic Rays}}

For the cold, nonrelativistic, $T_{cr}\ll m_{cr}c^{2}$, cosmic rays, the
linearized Equation (6) takes the form%
\begin{equation}
\gamma_{cr0}\frac{\partial\mathbf{v}_{cr1}}{\partial t}+\gamma_{cr0}^{3}%
\frac{\mathbf{u}_{cr}}{c^{2}}\mathbf{u}_{cr}\cdot\frac{\partial\mathbf{v}%
_{cr1}}{\partial t}=-\frac{\mathbf{\nabla}p_{cr1}}{m_{cr}n_{cr0}}%
+\mathbf{F}_{cr1}\mathbf{+}\frac{q_{cr}}{m_{cr}c}\mathbf{v}_{cr1}%
\times\mathbf{B}_{0},\tag{B1}%
\end{equation}
where
\begin{equation}
\mathbf{F}_{cr1}=\frac{q_{cr}}{m_{cr}}\left(  \mathbf{E}_{1}\mathbf{+}\frac
{1}{c}\mathbf{u}_{cr}\times\mathbf{B}_{1}\right) \tag{B2}%
\end{equation}
and $\mathbf{u}_{cr}$ is directed along the $y$-axis. We have used that
$\gamma_{cr1}=\gamma_{cr0}^{3}\mathbf{u}_{cr}\cdot\mathbf{v}_{cr1}/c^{2}$,
where $\gamma_{cr0}=\left(  1-u_{cr}^{2}/c^{2}\right)  ^{-1/2}$. Equations
(B1) and (B2) do not include $\mathbf{v}_{e0}$ in Equation (13). From Equation
(B1), we find the following equations for $v_{cr1x,y}$:%
\begin{align}
\left(  \gamma_{cr0}^{4}\frac{\partial^{2}}{\partial t^{2}}\mathbf{+}%
\omega_{ccr}^{2}\right)  v_{cr1x}  & =\omega_{ccr}F_{cr1y}+\gamma_{cr0}%
^{3}\frac{\partial F_{cr1x}}{\partial t},\tag{B3}\\
\left(  \gamma_{cr0}^{4}\frac{\partial^{2}}{\partial t^{2}}\mathbf{+}%
\omega_{ccr}^{2}\right)  v_{cr1y}  & =-\omega_{ccr}F_{cr1x}+\gamma_{cr0}%
\frac{\partial F_{cr1y}}{\partial t}.\nonumber
\end{align}
The $z$-component of Equation (B1) is given by
\begin{equation}
\gamma_{cr0}\frac{\partial v_{cr1z}}{\partial t}=-\frac{1}{m_{cr}n_{cr0}}%
\frac{\partial p_{cr1}}{\partial z}+F_{cr1z}\mathbf{.}\tag{B4}%
\end{equation}

From Equation (7) in the linear approximation, we find
\begin{equation}
p_{cr1}=p_{cr0}\Gamma_{cr}\left(  \frac{n_{cr1}}{n_{cr0}}-\gamma_{cr0}%
^{2}\frac{u_{cr}v_{cr1y}}{c^{2}}\right)  .\tag{B5}%
\end{equation}
Applying to Equation (B4) operator $\partial/\partial t$, substituting
Equation (B5), and using the continuity equation for cosmic rays, we obtain%
\begin{equation}
L_{cr}v_{cr1z}=H_{cr1}.\tag{B6}%
\end{equation}
Here,
\begin{align}
L_{cr}  & =\gamma_{cr0}\frac{\partial^{2}}{\partial t^{2}}-c_{scr}^{2}%
\frac{\partial^{2}}{\partial z^{2}},\tag{B7}\\
H_{cr1}  & =c_{scr}^{2}\gamma_{cr0}^{2}\frac{u_{cr}}{c^{2}}\frac{\partial
^{2}v_{cr1y}}{\partial z\partial t}+\frac{\partial F_{cr1z}}{\partial
t},\nonumber
\end{align}
where $c_{scr}=\left(  p_{cr0}\Gamma_{cr}/m_{cr}n_{cr0}\right)  ^{1/2}$ is the
sound speed of cosmic rays. We note that the first term on the right-hand side
in the definition of $H_{cr1}$ in Equation (B7) is connected with the
perturbation of the Lorentz factor.

\bigskip

\end{appendix}
\end{document}